\documentclass{article}[12pt]
\usepackage{epsf}
\usepackage[latin1]{inputenc}
\usepackage{amsmath,amsfonts,amssymb}
\usepackage{graphicx}

\def\beq{\begin{equation}}
\def\eeq{\end{equation}}
\def\bea{\begin{eqnarray}}
\def\eea{\end{eqnarray}}

\def\neq{\not=}

\date{}

\begin{document}

\begin{titlepage}
\begin{center}
{\large\bf
Spin-Singlet Quantum Hall States and Jack Polynomials with a Prescribed Symmetry
}\\[.3in] 

{\bf Benoit Estienne$^{1}$ and B.~Andrei Bernevig$^{2}$}\\
	$^1$ {\it Institute for Theoretical Physics, Universiteit van Amsterdam \\
Valckenierstraat 65, 1018 XE Amsterdam, The Netherlands\\
            e-mail: {\tt b.d.a.estienne@uva.nl}}\\
         
          $^2$  Department of Physics, Princeton University \\ Princeton, NJ 08544, USA \\ e-mail: {\tt bernevig@princeton.edu}\\

\end{center}
\centerline{(Dated: \today)}
\vskip .2in
\centerline{\bf ABSTRACT}
\begin{quotation}
We show that a large class of bosonic spin-singlet Fractional Quantum Hall model wavefunctions and their quasi-hole excitations can be written in terms of Jack polynomials with a prescribed symmetry. Our approach describes new spin-singlet quantum Hall states at filling fraction $\nu = \frac{2k}{2r-1}$ and  generalizes the $(k,r)$ spin-polarized Jack polynomial states. The NASS and Halperin spin singlet states emerge as specific cases of our construction. The polynomials express many-body states which contain configurations obtained from a root partition through a generalized squeezing procedure involving spin and orbital degrees of freedom. The corresponding generalized Pauli principle for root partitions is obtained, allowing for counting of the  quasihole states. We also extract the central charge and quasihole scaling dimension, and propose a conjecture for the underlying CFT of the $(k,r)$ spin-singlet Jack states. 
\end{quotation}
\end{titlepage}

\vskip 0.5cm
\noindent
{PACS numbers: 75.50.Lk, 05.50.+q, 64.60.Fr}

\section{Introduction}

Our understanding of the Fractional Quantum Hall effect has benefted tremendously from the existence of model trial wavefunctions against which the ground-state of a many-body realistic Coulomb Hamiltonian can be compared.  These model wavefunctions represent quantum amplitudes for the ground-state and excitations of many-electron systems in a magnetic field at rational filling factors. From a theoretical perspective, they allow for the determination of universal properties such as Hall conductance,  quantum numbers of the excitations, and more importantly, braiding statistics upon adiabatic exchange of excitations.

Understanding the structure of these model wavefunctions has been an important research topic in the past two decades. Despite having an explicit interacting wavefunction (such as the Laughlin state, which has a straight-forward from in real space) computation of important quantities such as correlation functions directly from the wavefunction has been elusive due to the extremely poorly understood expansion of these states in second-quantized basis. In \cite{Bernevig_Haldane1} it has been realized that many  (bosonic) Fractional Quantum Hall (FQH) wavefunctions, such as Laughlin \cite{Laughlin}, Moore-Read \cite{Moore_Read}, and Read-Rezayi \cite{Read_Rezayi}, as well as others, such as the state called the "Gaffnian"  \cite{Gaffnian},  could be be explicitly written as single Jack symmetric polynomials. The Jack polynomials have known expansions in terms of the second quantized basis of particles in the Lowest Landau Level (LLL), thereby solving one of the main difficulties of the expansion of the interacting state.   All the Jack spin polarized states have  enjoy clustering properties: they vanish with some power $r$ when $k+1$ particles come together. They are indexed by a "root partition", a specific configuration of the momenta of each of the electrons, which satisfied a generalized Pauli principle of not having mode than  $k$  particle in $r$ consecutive orbitals. This principle allows for the unification of a large class of FQH states. It also allows for generating them numerically much more efficiently than previously possible, as the Hilbert space dimension of the Jack polynomial with root partitions satisfying the generalized Pauli principle is small subset of the overall Hilbert space.

While spin-polarized wavefunctions are dominant in the study of FQH, spin-singlet wavefunctions are crucial at describing spin-unpolarized systems, bilayers or systems with valley degeneracy such as AlGaAs or graphene. Unfortunately, numerical studies of spin-singlet systems are hampered by the exponential growth of the Hilbert space, which is much more severe than in the spin-polarized case. For this reason, a method that allows for the determination of the Hilbert space configurations and their weights is necessary.  In this paper we take an important step and  extend the Jack polynomial  approach to a large class of spin singlet FQH wavefunctions. We introduce the non-symmetric Jack polynomials, define a Pauli principle for their root partitions, and show that they satisfy  clustering conditions similar to those of their spin-polarized counterparts.  The Halperin and NASS states  emerge as special cases of our construction. We identify a spin-Laplace Beltrami operator that diagonalizes our states, and end our paper by presenting and substantiating  a conjecture relating the non-symmetric Jack polynomials to specific $\mathcal{W}$-conformal field theory (CFT) models.

\section{Spinfull FQH states and Spin Calogero Sutherland}

In this paper we are concerned with particles with spin having two internal states $\sigma \in \{ \uparrow, \downarrow\}$. For most of the paper we focus on this case, although in the last section we relax this constraint and consider particles with an arbitrary number $n$ of internal states. We start by fixing the notation conventions.
For spinfull particles in the Lowest Landau level,  manybody wavefunctions (for $N$ particles) are of the form:
\begin{equation}
\Psi(z_1,\cdots, z_{N}) = \sum_{\{\sigma_i\}} \psi(z_i,\sigma_i) | \sigma_1 \cdots \sigma_{N} \rangle  \label{spin_wf}
\end{equation}
where the coordinates $z_i$ live in the complex plane, and $ \psi(z_i,\sigma_i)$ are polynomials is $z_i$. For indiscernible particles such a wavefunction must be completely (anti)symmetric under simultaneous exchange of  position and spin:
 \begin{equation}
K_{ij} P_{ij} \Psi(z_1,\cdots, z_{N}) = \epsilon \Psi(z_1,\cdots, z_{N})  \qquad \epsilon = \pm 1 \label{symmetry}
\end{equation}
where we introduced two exchange operators $K_{ij}$ and $P_{ij}$. The first one exchanges the positions of particles $i$ and $j$
\begin{equation}
K_{ij} z_i = z_j K_{ij}  \label{K_ij}
\end{equation}
while $P_{ij}$ exchanges the spins of particles $i$ and $j$. 
\begin{equation}
P_{ij}|  \cdots \sigma_i \cdots \sigma_j \cdots \rangle =  |  \cdots \sigma_j \cdots \sigma_i \cdots \rangle \label{P_ij}
\end{equation} 
Whenever the quantum system under study enjoys a spin $\textrm{SU}(2)$ symmetry (or a broken symmetry preserving $S_z$, for instance in the presence of  Zeeman interaction), $S_z$ is a good quantum number, and all eigenstates of the Hamiltonian can be chosen to have a well defined number $N_{\uparrow}$ of up-spins, and $N_{\downarrow} = N - N_{\uparrow}$ of  down-spins.

In principle,  all $2^N$ components $\{ \psi(z_1,\cdots, z_N |\sigma) ,  \sigma \in \{ \uparrow, \downarrow\}^N \}$ are necessary to describe the wavefunction \eqref{spin_wf} . However using the permutation symmetry \eqref{symmetry} and working at constant $S_z = \frac{1}{2} \left( N_{\uparrow} - N_{\downarrow}\right) $ simplifies the situation drastically. In this case a  single component is sufficient to store all the information about the wavefunction:
\begin{equation}
\Psi(z_1,\cdots, z_{N}) = \mathcal{S} \left( \Phi(z_1, \cdots, z_N) | \underbrace{\uparrow \cdots \uparrow}_{N_{\uparrow}} \underbrace{\downarrow \cdots \downarrow}_{N_{\downarrow}}  \rangle  \right) \label{S_Phi}
\end{equation}
where $ \mathcal{S}$ is a total (anti)symmetrization, acting on both positions and spins.  Without any loss of generality, the component $\Phi(z_1, \cdots, z_N)$ can be chosen to be  $\mathfrak{S}_{N_{\uparrow}}\otimes \mathfrak{S}_{N_{\downarrow}}$ (anti)symmetric. In the following we will sometimes refer to $\Phi$ as the wavefunction, but it has to be understood in the sense of \eqref{S_Phi}.

A convenient basis for the space of $\mathfrak{S}_{N_{\uparrow}}\otimes \mathfrak{S}_{N_{\downarrow}}$ (anti)symmetric polynomials is the set of monomials $m_{\lambda_{\uparrow},\lambda_{\downarrow}}$:
\begin{equation}
m_{\lambda_{\uparrow},\lambda_{\downarrow}} = m_{\lambda_1}(z_1, \cdots, z_{N_{\uparrow}}) m_{\lambda_2}(z_{N_{\uparrow}+1},\cdots, z_N)
\end{equation}
where $m_{\lambda}$ are the usual totally (anti)symmetric monomials. This basis is naturally indexed by two partitions $\lambda_{\uparrow}$ and $\lambda_{\downarrow}$ with $N_{\uparrow}$ and $N_{\downarrow}$ entries, respectively, which label the set of angular momenta of the particles. However for the purpose of the present article, it is more convenient to label this basis by a partition with $N$ entries $\lambda$ and a spin dressing $\sigma = | \sigma_1 \cdots \sigma_N \rangle$, defined by
\begin{equation}
\mathcal{S} \left( m_{\lambda}(z_1, \cdots, z_N) | \sigma_1 \cdots \sigma_N \rangle \right) =  \mathcal{S} \left( m_{\lambda_{\uparrow},\lambda_{\downarrow}}(z_1, \cdots, z_N) |\uparrow \cdots \uparrow \downarrow \cdots \downarrow \rangle \right)\label{lambda_sigma}
\end{equation}
We impose the rule that whenever $\lambda_i=\lambda_{i+1}$, we choose $\sigma_i \geq \sigma_{i+1}$, to avoid overcounting states. The mapping from the dressed partition $(\lambda,\sigma)$ notation to the more natural $(\lambda_{\uparrow},\lambda_{\downarrow})$ is simply the following:  one can reconstruct the two partitions $\lambda_{\uparrow}$ and $\lambda_{\downarrow}$ by partitioning $\lambda$ according to  $\lambda_i \in \lambda_{\uparrow}$ for $\sigma_{i} = \uparrow$ and $\lambda_i \in \lambda_{\downarrow}$  for $\sigma_{i} = \downarrow$. As we will see, such a spin-dressed partition $(\lambda, \sigma)$ allows for defining a Pauli principle that counts the excitations of many spin-singlet FQH states.

\subsection{Spin-singlet states with clustering properties}

Although the fractional quantum Hall effect happening in strong magnetic fields, it is well-known that some FQH ground states are spin-unpolarized. Spin-unpolarized model wavefunctions are also of use in systems with spin and valley degeneracy, such as bilayer quantum Hall states, graphene or the GaAs hole-band systems. From both  a theoretical and numerical point of view, the structure of interacting many-body spin-unpolarized states needs to  be better understood.   Like their spin-polarized counterparts, the spin singlet model FQH ground-states and their excitations can, in principle, be characterized by clustering conditions and contain hidden algebraic structure that we aim to uncover.

Wavefunctions for $N$ particles with spin in the Lowest Landau level (LLL) are of the generic form  \eqref{S_Phi}. In the context of the FQHE, it is conventional to denote by $z_i$ and $w_j$  the positions of the up-spins and down-spins particles, respectively. For instance the simplest spin-unpolarized wavefunctions, the $(r,r,n)$ Halperin ground state wavefunctions \cite{Halperin} are given by
\begin{equation}
\Phi^{(r,r,n)}(z_i, w_j) = \prod_{i<j} (z_i-z_j)^r \prod_{i<j} (w_i-w_j)^r \prod_{i,j} (z_i-w_j)^n \label{Halperin_wf}
\end{equation} 
As usual when dealing with particles in the LLL, we dropped the trivial gaussian factors $\exp \left( -\sum_i (|z_i|^2 +|w_i|^2)/4l^2 \right)$ with $l= \sqrt{\hbar/eB}$ the magnetic length. If the interactions between electrons preserve the spin symmetry, the Hilbert space can be decomposed into irreducible representations of $\textrm{SU}(2)$. In particular if the ground state is singly degenerate, it has to be a spin singlet ($S=0$).  For the Halperin wavefunction \eqref{Halperin_wf}, this only holds when $n=r-1$.  The spin-singlet Halperin state has filling fraction $\nu = 2/(2r-1)$, as can be seen from power counting in
\begin{equation}
\Phi^{(r)}_H(z_i, w_j) = \prod_{i<j} (z_i-z_j)^r \prod_{i<j} (w_i-w_j)^r \prod_{i,j} (z_i-w_j)^{r-1} \label{Halperin_SS_wf}
\end{equation} 
The Halperin wavefunction \eqref{Halperin_SS_wf} is in many ways the most natural spin-singlet extension of the spin-polarized Laughlin wavefunction
\begin{equation}
\Phi^{(r)}_L(z_i) = \prod_{i<j} (z_i-z_j)^r \label{Laughlin_wf}
\end{equation} 
They enjoy the same clustering properties, and they both support excitations that are fractional but abelian. For spin-polarized states, the Laughlin wavefunction is the foundation of a series of more complicated states, exhibiting non-abelian statistics, called the Read-Rezayi (RR) states. The bosonic RR spin polarized states are indexed by an integer $k$ and are formed by dividing the electrons into $k$ groups of $N/k$ particles each, forming $\nu=1/2$ filling Laughlin states  out of each $N/k$ electrons, multiplying the wavefunctions and then symmetrizing over the coordinates of the $k$ clusters.   In \cite{NASS1} a new-class of non-abelian spin-singlet (NASS) wavefunctions have been introduced. In the same sense that Halperin is an extension of Laughlin to spin-singlet states, these wavefunctions generalize the Moore-Read and Read-Rezayi states. They can be obtained as symmetrizations  of $k$ clusters of Halperin wavefunctions. For an extensive introduction to the NASS states we engage the reader to Ref.  \cite{Ardonne_thesis}. 

The Laughlin, Moore-Read and Read-Rezayi states can be seen as belonging to the same family of FQH states. A property of these wavefunctions vanish when $k+1$ particles come together, for the values of $k$ listed in Table \eqref{tab:clustering}. This also holds true for the Halperin and NASS wavefunctions. This makes them the unique ground states of  of Haldane-type pseudopotentials which disallow a cluster of particles to have angular momentum smaller than a specified value.

The Laughlin, Moore-Read and Read-Rezayi states belong to an even larger family of clustering states. A bosonic FQH state is said to enjoy $(k,r)$ clustering properties if all its zero energy wavefunctions vanish with a power $r$ when $k+1$ particles come together. As long as $k+1$ and $r-1$ are coprime, one can define a symmetric Jack polynomial at negative rational coupling $\alpha = - (k+1)/(r-1)$ satisfying this $(k,r)$ clustering properties, which can be in turn interpreted as a trial FQH wavefunction \cite{Bernevig_Haldane1}.  For $r=2$ one recovers Laughlin, Moore-Read and Read-Rezayi for $k=1,2$ and $k\geq 3$, respectively. The Gaffnian \cite{Gaffnian} also belongs to this family, and correspond to $(k,r)=(2,3)$. The Jack polynomial approach unifies these seemingly different states in a common framework and  allows for their efficient numerical generation.  In this paper we extend the Jack polynomial approach to spin-singlet states.

\begin{table*}[t]
 \begin{center}
\begin{tabular}{c|c|c|}
 & spin-polarized & spin-singlet \\
 \hline
$k=1$         & ($r=2$) Laughlin & ($r=2$) Halperin \\
$k=2$         & Moore-Read     & $k=2$ NASS\\
$k\geq 3$  & Read-Rezayi     &  $k \geq 3$ NASS\\
\hline
\end{tabular}
\end{center}
\caption{Classification of spin-polarized and spin-singlet clustering states enjoying ($r=2$) clustering properties.}
\label{tab:clustering}
\vspace{-5pt}
\end{table*}

\subsection{Spin-Calogero-Sutherland Model}

Symmetric Jack polynomials appear naturally in physics as eigenstates of the Calogero-Sutherland (CS) Hamiltonian. The CS Model \cite{Sutherland,Calogero} describes (spinless) particles on a circle interacting with a long range potential. The positions of the $N$ particles are denoted $x_i$, $0 \leq x_i \leq L$. The total momentum and the CS Hamiltonian are respectively given by
\begin{align}
\hat{P} & = \sum_{j=1}^N \frac{1}{i} \frac{\partial}{\partial x_j} \\
\hat{H} & = -\frac{1}{2} \sum_i\left( \frac{\partial}{\partial x_i} \right) ^2 + \left( \frac{\pi}{L} \right)^2 \sum_{i<j} \frac{\beta(\beta -1)}{\sin^2 (\pi (x_i-x_j)/L)} \label{calogerosutherland1}
\end{align}
where $\beta$ is a parameter that indexes the operator. In this paper we are interested in the spin generalization of the Calogero-Sutherland Model, introduced in \cite{Bernard_Gaudin_Haldane_Pasquier,Cherednik}. 
Each particle carries a spin with $2$ possible values, and the dynamics of the model are governed by the generalization of the  Hamiltonian in Eq[\ref{calogerosutherland1}]:
\begin{equation}
\hat{H} = - \frac{1}{2} \sum_i\left( \frac{\partial}{\partial x_i} \right) ^2 + \left( \frac{\pi}{L} \right)^2 \sum_{i<j} \frac{\beta(\beta -\epsilon P_{ij})}{\sin^2 (\pi (x_i-x_j)/L)}
\end{equation}
where $\epsilon = 1$ ($-1$) for bosons (fermions) and the permutation $P_{ij}$ is the one from \eqref{P_ij} and  permutes the spins.
From now on we shall work on the unit circle $L=2\pi$ and in the complex variables $z_j = \exp ( i  x_j)$ :
\begin{equation}
\hat{H} =  \sum_i\left( z_i\frac{\partial}{ \partial z_i} \right) ^2 - \sum_{i\neq j} \beta(\beta -\epsilon P_{ij})\frac{z_iz_j}{(z_i-z_j)^2}
\end{equation}
The eigenstates of $\hat{H}$ have the following structure:
\begin{equation}
\psi(z_i,\sigma_i)  \prod_{i<j} (z_i-z_j)^{\beta}
\end{equation}
where the wave function $\Phi(z_i,\sigma_i)$ is completely (anti)symmetric under the simultaneous permutations of the spin and the coordinates. It is convenient to work with the effective Hamiltonian $H$ acting on $\psi(z_i,\sigma_i)$:
\begin{equation}
H = \sum_i\left( z_i\frac{\partial}{ \partial z_i} \right) ^2 + \beta \sum_{i<j} \frac{z_i+z_j}{z_i-z_j}\left( z_i \frac{\partial}{\partial z_i} - z_j \frac{\partial}{\partial z_j} \right) -\beta \sum_{i\neq j}(1-  \epsilon P_{ij})  \frac{z_iz_j}{(z_i-z_j)^2}
\end{equation}
and we rename $\beta \to 1/\alpha$. If the last term in the above operator were missing, this would be the usual Laplace-Beltrami operator. Since the total spin and coordinate wave function is (anti)symmetric, it satisfies:
\begin{equation}
K_{ij} \Phi = \epsilon P_{ij} \Phi
\end{equation}
where $K_{ij}$ is the exchange operator defined in \eqref{K_ij}.  One ends up with the following Hamiltonian:
\begin{equation}
H_{sLB}^{(\alpha)} = \sum_i \left(z_i \frac{\partial}{\partial z_i} \right)^2 + \frac{1}{\alpha} \sum_{i<j} \frac{z_i+z_j}{z_i-z_j} \left( z_i \frac{\partial}{\partial z_i} - z_j \frac{\partial}{\partial z_j} \right) - \frac{1}{\alpha}  \sum_{i\neq j}  \left(1-K_{ij}\right) \frac{z_i z_j}{(z_i-z_j)^2} \label{sLB}
\end{equation}
This operator differs from the usual Laplace-Beltrami (LB) operator only in the presence of the extra last term involving the exchange operators $K_{ij}$. If the Hilbert space under consideration is that of symmetric functions, the exchange operators act trivially ($K_{ij}=1$) and one recovers the usual LB operator. For this reason we denote the Hamiltonian \eqref{sLB} spin-Laplace-Beltrami (sLB) operator. 

When acting on the larger space of non-symmetric functions, the usual LB operator does not preserve the subspace of polynomials. Moreover it is no longer integrable, as the very nature of the Dunkl operators underlying the integrability  structure requires the presence of theses exchange operators \cite{Bernard_Gaudin_Haldane_Pasquier}. For both these reasons, the sLB is the correct operator to consider if one is interested in non-symmetric polynomials. Not only is the sLB integrable even when acting on non-symmetric polynomials, but it also preserves any subspace of polynomials with an arbitrary prescribed symmetry, for instance $\mathfrak{S}_{N_{\uparrow}}\otimes \mathfrak{S}_{N_{\downarrow}}$. As a side remark, the sLB operator is also the correct approach to deal with spin-polarized antisymmetric  wavefunctions. As was noted in \cite{Product_rule1}, the Laplace Beltrami operator needs to be modified for fermions, and this is precisely taken care of by the exchange term in \eqref{sLB}.  

\subsection{Squeezing for Jack polynomials with a prescribed symmetry}

The non-symmetric Jack polynomials are eigenfunctions of the spin-Laplace-Beltrami operator \eqref{sLB}.  In the appropriate basis the sLB operator is triangular \cite{Bernard_Gaudin_Haldane_Pasquier}, and this implies a notion of squeezing for non-symmetric Jack polynomials. Using a partial symmetrization of these non-symmetric Jack polynomials, it is possible to construct eigenstates of the sLB operator with a prescribed symmetry \cite{Baker_Forrester}. In particular for $\mathfrak{S}_{N_{\uparrow}}\otimes \mathfrak{S}_{N_{\downarrow}}$ symmetric polynomials, the induced ordering for dressed partitions $(\lambda,\sigma)$ is:
\begin{equation}
(\lambda,\sigma) >  (\mu,\sigma') \Leftrightarrow \lambda > \mu, \textrm{ or } \lambda = \mu \textrm{ and } \sigma>\sigma'
\end{equation}
  $\lambda> \mu$ is the usual dominance partial ordering of partitions, and $\sigma>\sigma'$ means $|\sigma| = |\sigma'|$ and lexicographic order ($\sigma_1 > \sigma_1' $ or $\sigma_1 =\sigma_1' $ and $\sigma_2 > \sigma_2' $  etc ). For instance $(4,1),(\uparrow,\downarrow) > (3,2),(\uparrow,\downarrow)  > (3,2),(\downarrow,\uparrow)$. We then say that $(\lambda, \sigma)$ dominates $(\mu, \sigma')$ or, alternatively, $(\mu, \sigma')$ is squeezed from $(\lambda, \sigma)$. This procedure generalizes the usual squeezing operation presented in \cite{Bernevig_Haldane1}.  With this ordering of dressed partitions,  a $\mathfrak{S}_{N_{\uparrow}}\otimes \mathfrak{S}_{N_{\downarrow}}$ symmetric Jack polynomial $J^{\alpha}_{(\lambda,\sigma)}$ can be expanded as
\begin{equation}
J^{\alpha}_{(\lambda,\sigma)}=m_{(\lambda,\sigma)}+\sum_{(\mu,\sigma')< (\lambda,\sigma)} u_{(\lambda, \sigma), (\mu,\sigma')}(\alpha)m_{(\mu,\sigma')} \label{Jack_expansion}
\end{equation}
and it is an eigenstate of sLB($\alpha$) with an energy
\begin{align}
E_{(\lambda,\sigma)}(\alpha) = \sum_{i=1}^N \lambda_i \left( \lambda_i + \frac{1}{\alpha}(N+1-2i) \right) \label{energy}.
\end{align}
This eigenvalue is independent of the spin-dressing $\sigma$ simply because the sLB operator commutes with the group of permutations $\mathfrak{S}_N$, and this explains the degeneracies in its spectrum.

In the orbital occupation number notation we encode a state in two sequences of numbers: the occupation numbers for up-spins $n^{\uparrow} = [n_0^{\uparrow},n_1^{\uparrow},\cdots]$ and for down-spins $n^{\downarrow} = [n_0^{\downarrow},n_1^{\downarrow},\cdots]$. In this language squeezing consists of two possible moves:
\begin{itemize}
\item a  spin-blind move: a usual squeeze move on the total occupancy $n=n^{\uparrow}+n^{\downarrow}$, followed by an \underline{arbitrary} spin dressing
\item a spin move, consisting in exchanging an up-spin and a down-spin, with the only constraint that the up-spin must be initially located on a higher orbital than the down spin (i.e. on its right in usual FQH conventions).
\end{itemize}
Note that this process does not treat up- and down-spins  equally. This is the price we pay in order to have a unique root partition. If both spins are treated equally, we cannot, in general, define a unique root partition and a squeezing rule. Unfortunately, not treating the two spins equally  means that our squeezing rule misses some constraints that could have been otherwise obtained.
For instance the the $r=2$ Halperin wavefunction with $N=4$ particles has root occupancy $(\downarrow,\uparrow,0,\downarrow,\uparrow)$. Squeezing from this gives the following states:
 \begin{align}
 & (\uparrow,\downarrow,0,\downarrow,\uparrow) & (\downarrow,\uparrow,0,\uparrow,\downarrow) & & (\uparrow,\downarrow,0,\uparrow,\downarrow) &&(\uparrow,\uparrow,0,\downarrow,\downarrow)
 \end{align}
 Our squeezing rule would imply that all the coefficients of the above partitions \emph{may} be nonzero.  However, a simple reflection shows us that the coefficient of $(\uparrow,\uparrow,0,\downarrow,\downarrow)$ must vanish: if we treated the spins on the same footing, exchanging the role of up and down spins leads us to the partition $(\downarrow,\downarrow, \uparrow \uparrow)$, which by spin symmetry has identical weight in a spin-singlet ground-state. However,  $(\downarrow,\downarrow, \uparrow \uparrow)$ is not squeezed from $(\downarrow,\uparrow,0,\downarrow,\uparrow)$ so it must have vanishing weight. However, we find this inconvenience to be much less bothersome than the use of multiple root partitions.

An important remark is that breaking down the permutation symmetry from $\mathfrak{S}_N$ to $\mathfrak{S}_{N_{\uparrow}}\otimes \mathfrak{S}_{N_{\downarrow}}$ introduced huge degenracies in the spectrum of the sLB operator. Eigenstates belong to multiplets of the remaining group $\mathfrak{S}_N / (\mathfrak{S}_{N_{\uparrow}}\otimes \mathfrak{S}_{N_{\downarrow}})$, and their degeneracy are typically of order $\binom{N}{N_{\uparrow}}$. For numerical applications it means that the sLB operator is not sufficient to obtain recursion relations \cite{Product_rule1,Product_rule2} for the non-symmetric Jack polynomials. For the FQH ground state however, the wavefunction is a spin-singlet. It is possible (for the cases we checked) that this extra constraint lifts the degeneracy. We plan to address this question, together with the numerical implementation of the recursion relations for non-symmetric Jack polynomials, in a forthcoming publication.

\section{Spin-singlet states as eigenstates of spin Laplace Beltrami}

In this section we define a new class of bosonic spin-singlet wavefunctions with $(k,r)$ clustering properties using the theory of non-symmetric Jack polynomials. For $k=1$ this is simply the Halperin wavefunction, while for $k\geq 2$ and $r=2$ the NASS state is recovered. We first give the "generalized Pauli principle" corresponding to this $(k,r)$ clustering. We then check our approach by proving that the $r$ Halperin and NASS ground state wavefunctions are eigenstates of the sLB operator. We then use this principle to recover the zero-mode counting of the $k=2$ NASS state, and we give the counting of the spinfull equivalent of the Gaffnian state $(k=2,r=3)$.

\subsection{Generalized exclusion principle, $(k,r)$ admissibility}
\label{Counting}

Jack polynomials, both symmetric \cite{FJMM1,FJMM2} and non-symmetric \cite{Kasatani}, enjoy $(k,r)$ clustering properties at the following negative value of the coupling 
\begin{equation}
\alpha = - \frac{k+1}{r-1}
\end{equation}
As for the symmetric case, non-symmetric Jack polynomials are usually well defined for positive values of the coupling $\alpha$, and an admissibility condition (or "generalized Pauli principle" in the language of \cite{Bernevig_Haldane2}) must be introduced to insure that a negative coupling does not introduce unbounded functions (that the weight of all configurations is finite). In particular for $\mathfrak{S}_{N_{\uparrow}}\otimes \mathfrak{S}_{N_{\downarrow}}$ symmetric Jack polynomials, a dressed partition $(\lambda,\sigma)$ is called "$(k,r)-$admissible" (for $k+1$ and $r-1$ coprime) if it obeys 
\begin{align}
& \lambda_{i}-\lambda_{i+k} \geq r-1, & \lambda_{i}-\lambda_{i+k} = r-1 \, \Rightarrow \, \left(\sigma_i,\sigma_{i+k}\right) = (\uparrow,  \downarrow) \label{spin_admissibility}
\end{align}
The fully polarized "generalized Pauli principle" \cite{Bernevig_Haldane2} is recovered when all particles have the same spin. This was expected as the fully polarized $(k,r)$ wavefunctions can be thought of as quasihole of the spin-singlet ones.
For completeness we mention the $(k,r)$ admissibility for partitions dressed by a $n$ states spin texture ($\sigma_i \in \{ 1,2,\cdots,n \}$):
\begin{align}
& \lambda_{i}-\lambda_{i+k} \geq r-1, & \lambda_{i}-\lambda_{i+k} = r-1 \, \Rightarrow \, \sigma_i < \sigma_{i+k} \label{spin_n_admissibility}
\end{align}
although in the following we focus on $n=2$. 

For $(k,r)$-admissible partitions $\mathfrak{S}_{N_{\uparrow}}\otimes \mathfrak{S}_{N_{\downarrow}}$ symmetric Jack polynomials are well defined at $\alpha = -\frac{k+1}{r-1}$, and enjoy the clustering properties inherited from the wheel condition of \cite{Kasatani,Kasatani_Pasquier}. In particular
\begin{align}
P(z_1,\cdots, z_N) = \prod_{a=k+1}^N (z_{i_a}-Z)^{r-1} Q(z_1,\cdots,z_N),    & & z_{i_a} = Z, \, a=1,\cdots,k
\end{align}
For $r=2$  wavefunctions satisfying this clustering are the unique exact zero-energy eigenstates of certain k + 1-body interaction Hamiltonian, in a similar way as the spin-polarized cases. However for $r >2$ this is usually not sufficient to characterize the FHQ state  (same situation occurs in the spin-polarized state \cite{Estienne_Regnault_Santachiara}) and for these cases we do not know a local Hamiltonian having the $(k,r)$ Jack non-symmetric polynomials as unique zero-modes (the Laplace Berltrami operator is nonlocal). As soon as  $r\geq3$ such a local Hamiltonian for the polarized case is only known for the Gaffnian. It is quite possible that a similar construction would work for the Gaffnian with $n$ internal states.

Nonetheless, the set of all admissible Jack polynomials with a prescribed symmetry $\mathfrak{S}_{N_{\uparrow}}\otimes \mathfrak{S}_{N_{\downarrow}}$, for any number of up-spins and down-particles, describes the set of all "zero energy" modes of this FQH state. Counting these states boils down to counting admissible partitions, which we illustrate in section \ref{counting}.

 Moreover this set of wavefunctions (in the sense of \eqref{S_Phi}) is stable under the action of the total $\textrm{SU}(2)$ spin generators $S^{\pm} = \sum_i S_{i}^{\pm}$ and $S^z = \sum_i S_{i}^{z}$.  This property is inherited from the stability of  the ideal of non-symmetric Jack polynomials at $\alpha = - (k+1)/(r-1)$ with $(k,r)$ admissible partitions under the full group of permutations $\mathfrak{S}_N$ \cite{Kasatani}. 

Finally we claim that this set of non-symmetric Jack polynomial is also stable under the action of the total angular momentum operators $L^{\pm} = \sum_i L_{i}^{\pm}$ and $L^z = \sum_i L_{i}^{z}$. This conjecture is supported by the underlying CFT of section \ref{CFT}. Indeed, this stability is a consequence of global conformal transformations (generated by the Virasoro modes $L_0,L_{\pm 1}$) of the underlying CFT, as was observed in the symmetric case \cite{EPSS}.

The densest $(k,r)-$admissible partition is unique, and reads in orbital occupation as:
\begin{equation}
\left(k\downarrow,0^{r-2},k\uparrow,0^{r-1},k\downarrow,0^{r-2},k\uparrow,0^{r-1}, \cdots, k \downarrow,0^{r-2},k\uparrow  \right)
\end{equation}
and the corresponding state lives on a sphere pierced by $N_{\Phi}$ flux quanta, where 
 \begin{equation}
 N_{\phi} = \frac{2r-1}{2k} N - r.
 \end{equation} 
 The unicity of the densest root partition, together with the stability of the set of Jack polynomials under $\vec{L}$ and $\vec{S}$, implies that the corresponding "ground-state" is rotationally invariant  on the sphere ($L=0$), as it should. Moreover it is a spin-singlet ($S=0$).  

In summary, these $(k,r)$ Jack polynomials with a prescribed symmetry describe a spin-singlet state at filling fraction $\nu = 2k/(2r-1)$.

We also mention the generalization of the root partition to the case of $n$ internal states:
\begin{equation}
\left(k \, (n),0^{r-2},k\, (n-1),\cdots,0^{r-2}, k\, (1),0^{r-1}, k \, (n),0^{r-2},k\, (n-1),\cdots,0^{r-2}, k\, (1), \cdots  \right)
\end{equation}
with filling fraction $\nu = nk / (n(r-1)+1)$.

\subsection{Spin-singlet Halperin state}

As has been done for the spin-polarized Laughlin \cite{Bernevig_Haldane1}, it is rather straightforward (see Appendix \ref{Halperin_Proof}) to check that the bosonic spin-singlet Halperin ground state wavefunction \eqref{Halperin_SS_wf}. 
\begin{equation}
\Phi^{(r)}_H(z_1, \cdots,z_{N/2} ,w_1,\cdots, w_{N/2}) = \prod_{i<j} (z_i-z_j)^r (w_i -w_j)^r \prod_{i,j} (z_i-w_j)^{r-1}
\end{equation}
is an eigenstate of the sLB operator for the coupling constant $\alpha = -2/(r-1)$. Moreover the eigenvalue is of the form \eqref{energy} corresponding to the root partition (in orbital occupation):
\begin{equation}
\left(\downarrow,0^{r-2},\uparrow,0^{r-1},\downarrow,0^{r-2},\uparrow,0^{r-1}, \cdots,  \downarrow,0^{r-2},\uparrow  \right)
\end{equation}
where the spin dressing can be extracted explicitly from the wavefunction. One cannot help but notice that the spin-polarized Laughlin wavefunctions  are eigenstates of the (spin) Laplace-Beltrami operator for precisely the same value of $\alpha$.  This is no accident, since these wavefunctions describe the spin-polarized quasi-hole states of Halperin. It is very natural to conjecture that all quasi-hole wavefunctions of Halperin are described by eigenstates of the sLB at this negative coupling  $\alpha = -2/(r-1)$. Note that we had to restrict ourself to the bosonic case since $k+1=2$ and $r-1$ must be coprime for these Jack polynomials to be well defined.

\subsection{NASS state}

At this state we do not have a computational proof that $k$ NASS wavefunctions are eigenstates of the sLB operator for $\alpha = - (k+1)$. However it is known \cite{Kasatani} that non-symmetric Jack polynomials at negative coupling $\alpha = - (k+1)$ enjoy $(k,r=2)$ clustering properties, and this makes these polynomials zero-energy modes of the NASS Hamiltonian. For a flux  $N_{\Phi} = 3N/(2k)-2$, the NASS ground state on the sphere is unique. It turns out that at this flux , there is only one root occupation compatible with the generalized Pauli principle \eqref{spin_admissibility}, \begin{equation}
(k\downarrow,k\uparrow,0,k\downarrow,k\uparrow,0, \cdots k\downarrow,k\uparrow)
\end{equation}
so one can already conclude that the $k$ NASS ground state wavefunction is equal to this specific Jack polynomial. We checked explicitely that it holds in the case of $k=2$ NASS, for the $N=4$ and $N=8$ particles ground state. We found respectively:
\begin{equation}
\mathcal{H}_{SCS}^{(\alpha)} \Phi_{NASS}^{(k=2)}(z_1,z_2|w_1,w_2) = \left( 2+ \frac{4}{\alpha}\right) \Phi_{NASS}^{(k=2)} (z_1,z_2|w_1,w_2)
\end{equation}
for any value of $\alpha$, and
 \begin{equation}
\mathcal{H}_{SCS}^{(\alpha = -3)} \Phi_{NASS}^{(k=2)}(z_1,z_2,z_3,z_4|w_1,w_2,w_3,w_4) = \frac{100}{3} \Phi_{NASS}^{(k=2)} (z_1,z_2,z_3,z_4|w_1,w_2,w_3,w_4)
\end{equation}
which holds only for the expected $\alpha = -3$. Moreover in both cases the "energy" (Laplace-Beltrami eigenvalue) matches with that of the root partition
\begin{equation}
(2\downarrow,2\uparrow,0,2\downarrow,2\uparrow,0, \cdots 2\downarrow,2\uparrow)
\end{equation}
in Eq[\ref{energy}].

\subsection{Counting zero modes}
\label{counting}

The admissibility condition \eqref{spin_admissibility} is a powerful tool to count quasihole states. This gives a further check of our approach, as counting formulas are available for both NASS and Halperin quasiholes. As a non trivial check, we reproduced table II of \cite{NASS2} simply by counting  $(k,r)$ admissible partitions \eqref{spin_admissibility}. As a further illustration, we give the corresponding table for the "spin-Gaffnian" state, corresponding to $k=2,r=3$. 

\begin{itemize}

\item \underline{$N=4$, $\Delta N_{\Phi}=1$ }

\begin{center}
\begin{tabular}{c|cccc}
 \# = 20 & $S$ = & $0$ & $1$ & $2$ \\
 \hline
 $L = 0$ & & 1 & 0 & 1 \\ 
 $L = 1$ & & 0 & 1 & 0 \\ 
 $L = 2$ & & 1 & 0 & 0 
\end{tabular}
\end{center}

\item \underline{$N=4$, $\Delta N_{\Phi}=2$ }

\begin{center}
\begin{tabular}{c|cccc}
  \# = 105 & $S$ = & $0$ & $1$ & $2$ \\
 \hline
 $L = 0$ & & 2 & 0 & 1 \\ 
 $L = 1$ & & 0 & 2 & 0 \\ 
 $L = 2$ & & 2 & 1 & 1 \\
 $L = 3$ & & 0 & 1 & 0 \\
 $L = 4$ & & 1 & 0 & 0 
\end{tabular}
\end{center}

\item \underline{$N=4$, $\Delta N_{\Phi}=3$ }

\begin{center}
\begin{tabular}{c|cccc}
  \# = 335 & $S$ = & $0$ & $1$ & $2$ \\
 \hline
  $L = 0$ & & 2 & 0 & 1 \\
  $L = 1$ & & 0 & 3 & 0 \\
  $L = 2$ & & 3 & 2 & 2 \\
  $L = 3$ & & 1 & 3 & 0 \\
  $L = 4$ & & 2 & 1 & 1 \\
  $L = 5$ & & 0 & 1 & 0 \\
  $L = 6$ & & 1 & 0 & 0
\end{tabular}
\end{center}

\item \underline{$N=4$, $\Delta N_{\Phi}=4$ }

\begin{center}
\begin{tabular}{c|cccc}
  \# = 810 & $S$ = & $0$ & $1$ & $2$ \\
 \hline
  $L = 0$ & &  2 & 0 & 2 \\
  $L = 1$ & &  0 & 3 & 0 \\
  $L = 2$ & &  4 & 3 & 2 \\
  $L = 3$ & &  1 & 5 & 1 \\
  $L = 4$ & &  4 & 3 & 2 \\
  $L = 5$ & &  1 & 3 & 0 \\
  $L = 6$ & &  2 & 1 & 1 \\
  $L = 7$ & &  0 & 1 & 0 \\
  $L = 8$ & &  1 & 0 & 0
\end{tabular}
\end{center}

\item \underline{$N=8$, $\Delta N_{\Phi}=1$ }

\begin{center}
\begin{tabular}{c|cccc}
   \# = 105 & $S$ = & $0$ & $1$ & $2$ \\
 \hline
 $L = 0$ & & 2 & 0 & 1 \\ 
 $L = 1$ & & 0 & 2 & 0 \\ 
 $L = 2$ & & 2 & 1 & 1 \\
 $L = 3$ & & 0 & 1 & 0 \\
 $L = 4$ & & 1 & 0 & 0 
\end{tabular}
\end{center}

\item \underline{$N=8$, $\Delta N_{\Phi}=2$ }

\begin{center}
\begin{tabular}{c|cccccc}
   \# = 1764 & $S$ = & $0$ & $1$ & $2$  & $3$ & $4$ \\
 \hline
  $L = 0$ & & 5 & 1 & 3 & 0 & 1 \\
  $L = 1$ & &1 & 8 & 2 & 1 & 0 \\
  $L = 2$ & &8 & 7 & 6 & 1 & 0 \\
  $L = 3$ & &3 & 10 & 3 & 1 & 0 \\
 $L = 4$ & & 7 & 6 & 4 & 0 & 0 \\
 $L = 5$ & & 2 & 5 & 1 & 0 & 0 \\
 $L = 6$ & & 3 & 2 & 1 & 0 & 0 \\
 $L = 7$ & & 0 & 1 & 0 & 0 & 0 \\
 $L = 8$ & & 1 & 0 & 0 & 0 & 0
\end{tabular}
\end{center}

\end{itemize}

\section{Underlying Conformal Field Theory}
\label{CFT}

Being model, clustered states, it is very likely that the $(k,r)$ nonsymmetric Jack polynomials can be expressed as expectation values of primary field correlators in conformal field theory (CFT).  
 For the spin polarized case, such an analysis has already been performed. The appropriate CFT was conjectured to be a specific $\mathcal{W}$ algebra in \cite{FJMM1,Bernevig_Haldane3}, and this correspondence was finally proven in \cite{RB1,RB2}. In this section we try to identify the CFT underlying the  $(k,r)$ spin singlet states by extracting the central charge and quasihole dimensions directly from the wavefunctions and matching them with those of a known CFT.

\subsection{Extracting the central charge}

For $(k,r)$ symmetric Jack polynomials, the electron operator is a chiral field $\Psi_1$ generating a parafermionic algebra $\mathbb{Z}_k^{(r)}$, whose fusion rules are based on $\textrm{SU}(2)_k$ (i.e. Read-Rezayi states). Changing the value of $r$ only  modifies the conformal dimension of the parafermionic fields, but not their fusion rules.

\begin{figure}
 \centering
 \includegraphics[scale=0.8]{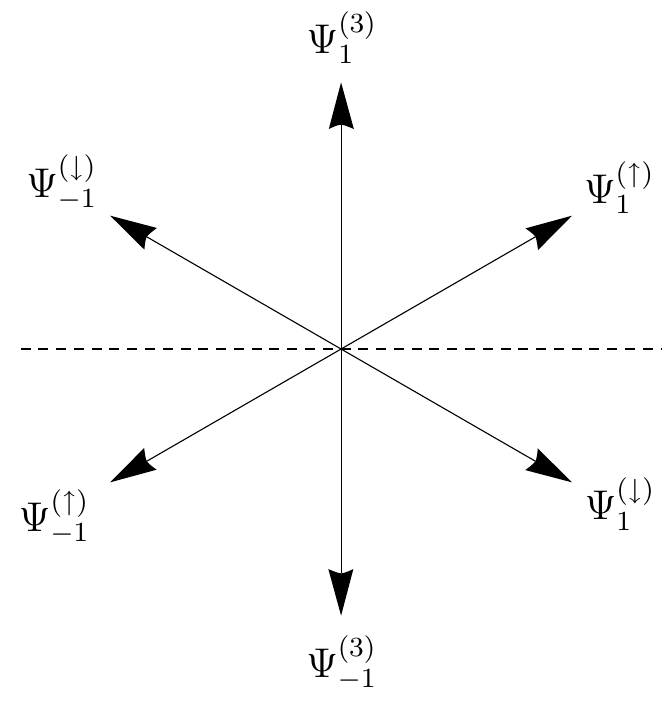}
 \label{SU(3)_lattice}
 \caption{Generators of the parafermionic algebra. This is symmetric under the reflexion around the dashed line, which implements the exchange of $\uparrow$ and $\downarrow$ required by $\textrm{SU}(2)$ symmetry of the FQH state.}
\end{figure}

For the spin-polarized case, there are two electron operators  $\Psi^{(\uparrow)} = \Psi_1^{(\uparrow)}$ and  $\Psi^{(\downarrow)} = \Psi_1^{(\downarrow)}$ whose fusion rules are dictated by those of the NASS state. More precisely the electron operators generates a parafermionic algebra of $\textrm{SU}(3)_k$ Gepner type \cite{Gepner}. This means that these fields live on the root lattice of $\textrm{SU}(3)_k$ (see Fig. \ref{SU(3)_lattice}).

We assume that one can write the non symmetric Jack polynomials as a conformal correlation function of the form
\begin{equation}
\langle \Psi^{(\uparrow)}(z_1) \dots \Psi^{(\uparrow)}(z_{N^{\uparrow}}) \Psi^{(\downarrow)}(w_1) \dots \Psi^{(\downarrow)}(w_{N^{\downarrow}}) \rangle \prod_{i<j} (z_{i}-z_j)^{\frac{r}{k}}  \prod_{i<j} (w_i-w_j)^{\frac{r}{k}}  \prod_{i,j} (z_{i}-w_j)^{\frac{r-1}{k}}  \label{CFT_wf}
\end{equation}

Their conformal dimensions depend on the integer $r$, in order to ensure the $(k,r)$ clustering properties of the polynomial \eqref{CFT_wf}.  As in the spin-polarized case, the fields $ \Psi_i^{(\uparrow)}$ and $ \Psi_i^{(\downarrow)}$ must have dimension $h_i = \frac{r}{2}\frac{i(k-i)}{k}$. But the conformal weights of all the other parafermionic fields, which can all be obtained by fusions of  the fundamental fields $ \Psi_i^{(\uparrow)}$ and $ \Psi_i^{(\downarrow)}$, are also determined by the clustering properties. For instance upon clustering $i$ up-spins and $k-i$ down-spins, the correlator \eqref{CFT_wf} must be non-singular. Moreover, this wavefunction must vanish with a power $r-1$ when another particle approaches this cluster. This forces the conformal dimension of the field obtained when fusing  $\Psi_i^{(\uparrow)}$  and  $\Psi^{(\downarrow)}_{k-i}$, which we call $\Psi^{(3)}_i$, to be $h^{(3)}_i = \frac{i(k-i)}{k}$.
\begin{align}
& \Psi_i^{(\uparrow)} \times \Psi^{(\downarrow)}_{-i} = \Psi^{(3)}_i                                             & h^{(3)}_i & = \frac{i(k-i)}{k} 
\end{align}
Demanding the wavefunction to be non-singular as we cluster one up-spin and one down-spin (or one up-spin and two down-spins), and non-vanishing when another particle approaches this cluster (for $k>2$  and $k>3$, respectively), we also found:
\begin{align}
& \Psi_i^{(\uparrow)} \times \Psi^{(\downarrow)}_{i} = \Psi^{(\uparrow \downarrow)}_i           & h^{(\uparrow \downarrow)}_1 & = (2r-1)\frac{(k-1)}{k} - (r-1) \\
& \Psi_i^{(\uparrow)} \times \Psi^{(3)}_{i} = \Psi^{(\uparrow 3)}_i           	 			   & h^{(\uparrow 3)}_1    & = \left( 2+\frac{r}{2}\right)\frac{(k-1)}{k} - 1
\end{align}

In order to extract the central charge of the CFT underlying the $(k,r)$ spin-singlet FQH state, one needs to write down the Operator Product Expansions (OPEs) up to level $2$ in the identity sector, and this is where the ambiguity lies. Usually one is led to assume that there is a single spin $2$ fields, namely the stress-energy tensor. But for the CFTs at work here, extra spin $2$ fields are required.  This is most easily seen from the characters of these CFTs, which can be obtained from the counting of section \ref{Counting}. Indeed, in the thermodynamic limit ($N,N_{\Phi} \rightarrow \infty$) the generating function counting the $(k,r)$ admissible partitions reduces to the character of the identity, up to two extra $\textrm{U}(1)$ factors coming from the two bosonic fields used to build the electron vertex operator. Removing theses bosonic degrees of freedom, one gets the following parafermionic character in the identity sector (for $k > 1$): 
\begin{equation}
\textrm{Tr} \left( q^{L_0}\right) = 1 + 3 q^2 + O(q^3)
\end{equation}
From this thermodynamic counting  of $(k,r)$ spin-admissible partitions there has to be two spin $2$ fields in the module of the identity besides the stress-energy tensor. We denote them by $W^{(\uparrow)}$ and $W^{(\downarrow)}$, and the OPEs can be put in the form
\begin{eqnarray}
\Psi^{(\uparrow)}_1(z)\Psi^{(\uparrow)}_{-1}(0) &= & z^{-r\frac{k-1}{k}}\left( 1 + \frac{r(k-1)}{kc}z^2 T(0) + z^2 C W^{(\uparrow)}(0)  + O(z^3) \right) \label{OPE1} \\
\Psi^{(\downarrow)}_1(z)\Psi^{(\downarrow)}_{-1}(0) &= & z^{-r\frac{k-1}{k}}\left( 1 + \frac{r(k-1)}{kc}z^2 T(0) + z^2 C W^{(\downarrow)}(0)  + O(z^3) \right) \label{OPE2}\\
\Psi^{(3)}_1(z)\Psi^{(3)}_{-1}(0) &= & z^{-2\frac{k-1}{k}}\left( 1 + \frac{2(k-1)}{kc}z^2 T(0) + z^2 D (W^{(\uparrow)}(0)+ W^{(\downarrow)}(0) ) + O(z^3) \right) \label{OPE3}
\end{eqnarray}
These are the most generic OPEs involving three spin $2$ fields,  after demanding invariance under the exchange of $\uparrow$ and $\downarrow$ required by $\textrm{SU}(2)$ symmetry of the spin-singlet FQH state. 

The final step is to compute some $4$ point functions. This is where we use the fact that any ground state wavefunction \eqref{CFT_wf} is by definition an eigenstate of  the spin CS Hamiltonian, corresponding to the densest admissible partition in the sense of \eqref{spin_admissibility}. We obtain (see Appendix \ref{hypergeometric}):
\begin{align}
 \langle \Psi^{(\uparrow)}_{-1}(0)\Psi^{(\uparrow)}_1(z)\Psi^{(\uparrow)}_{1}(1)\Psi^{(\uparrow)}_{-1}(\infty)\rangle z^{r\frac{(k-1)}{k}}(1-z)^{\frac{r}{k}}  & = {}_2F_{1}\left[-r,\frac{1-r}{k+1};k\frac{1-r}{k+1};z\right]  \label{hyper1} \\
 \langle \Psi^{(\uparrow)}_{-1}(0)\Psi^{(\uparrow)}_1(z)\Psi^{(\downarrow)}_{1}(1)\Psi^{(\downarrow)}_{-1}(\infty)\rangle z^{r\frac{(k-1)}{k}}(1-z)^{\frac{r-1}{k}}  &= {}_2F_{1}\left[1-r,\frac{1-r}{k+1};k\frac{1-r}{k+1};z\right] \label{hyper2}
\end{align}
Moreover the following two correlations functions are very strongly constrained by the low dimension of $\Psi^{(3)}_i$. They are polynomials with degree $1$ and $2$ respectively, and therefore the dominant terms in the OPEs as $z$ goes to $0$ and $\infty$ are sufficient to compute them:
\begin{align}
 \langle \Psi^{(3)}_{-1}(0)\Psi^{(3)}_1(z)\Psi^{(3)}_{1}(1)\Psi^{(3)}_{-1}(\infty)\rangle z^{2\frac{k-1}{k}}(1-z)^{\frac{2}{k}} & = 1 - \frac{2}{k}z + z^2 \\
 \langle \Psi^{(\uparrow)}_{-1}(0)\Psi^{(\uparrow)}_1(z)\Psi^{(3)}_{-1}(1)\Psi^{(3)}_{1}(\infty)\rangle z^{r\frac{k-1}{k}}(1-z)^{\frac{k-1}{k}} & = 1 - \frac{k-1}{k}z
 \end{align}

We are now in a position to extract the central charge $c$. Comparing the term $O(z^2)$ in these 4 correlators with the OPEs \eqref{OPE1}-\eqref{OPE3} one gets four equations, allowing to determine the four unknowns $c, C ,D$ and  $\gamma$ appearing in the OPEs. In particular one gets for the central charge:
\begin{equation}
c= -\frac{2 (k-1) (1+r((r-2)k-2))}{k+2 r-1}
\end{equation}
As a first check, Halperin $(k=1)$ has a trivial neutral CFT ($c=0$) as expected. Moreover one recovers the NASS central charge $c= 6(k-1)/(k+3)$ for $r=2$. These are the only cases when the underlying CFT is unitary, as for $r \geq 3$ the central charge is always negative (except $k=1$, i.e. Halperin). This mimics exactly the spin polarized case. In particular the "spin-Gaffnian" has central charge $c = -2/7$. The states with negative central charge are not expected to lead to a description of gapped topological phases \cite{Readnu1,Readnu2}.

\subsection{Underlying $\mathcal{W}$ algebra}

So far we have extracted the central charge of the CFT, and the first few terms of the character in the identity sector. It is very tempting to try to identify the underlying algebra. For the spin polarized case this was found to be a $\mathcal{W}$ algebra \cite{Bernevig_Haldane3}. This could be seen for Read-Rezayi states as coming from the coset equivalence
\begin{equation}
\frac{\textrm{SU}(2)_k}{\textrm{U}(1)} = \frac{\textrm{SU}(k)_1 \otimes \textrm{SU}(k)_1}{\textrm{SU}(k)_2}   \label{coset1}
\end{equation}
The l.h.s. is the parafermionic CFT $\mathbb{Z}_k$ responsible for the clustering properties of the RR states, while the r.h.s. is a specific $\mathcal{W}$ minimal model of the unitary series:
\begin{equation}
\frac{\textrm{SU}(k)_l \otimes \textrm{SU}(k)_1}{\textrm{SU}(k)_{l+1}}  \qquad l \geq 1\label{serie1}
\end{equation}
All these CFTs posses the same underlying $\textrm{WA}_{k-1}$ symmetry, and they are non unitary for a fractional level $l$ in \eqref{serie1}. The usual parametrization of the minimal models is $\textrm{WA}_{k-1}(p,q)$, with two coprime integers $(p,q)$ such that
\begin{align}
k+l & = \frac{p}{q-p} \\
k+l+1 & = \frac{q}{q-p} 
\end{align}
The corresponding central charge is
\begin{equation}
c_k(p,q) = (k-1)\left( 1- \frac{k(k+1)(p-q)^2}{pq} \right)
\end{equation}
For the NASS FQH state with $n$ internal states, the analog of the relation \eqref{coset1} can be found in \cite{diFrancesco}:
\begin{equation}
\frac{\textrm{SU}(n+1)_k}{\textrm{U}(1)} = \frac{(\textrm{SU}(k)_1)^{n+1}}{\textrm{SU}(k)_{n+1}}   \label{coset2}
\end{equation}
where $(\textrm{SU}(k)_1)^{n+1}$ stands for the direct sum of $n+1$ copies of $\textrm{SU}(k)_1$. A very natural guess for the $\mathcal{W}$ CFT is 
\begin{equation}
\frac{\textrm{SU}(k)_l \otimes (\textrm{SU}(k)_1)^{n}}{\textrm{SU}(k)_{n+l}}  
\end{equation}
Allowing the level $l$ to be fractional 
\begin{align}
k+l & = \frac{np}{q-p} \\
k+l+n & = \frac{nq}{q-p} 
\end{align}
leads to the following guess for the central charge of the minimal model $\mathcal{W}^{(n)}_k(p,q)$ 
\begin{equation}
c^{(n)}_k(p,q) = n(k-1)\left( 1- \frac{k(k+1)(p-q)^2}{n^2 pq} \right)
\end{equation}
To the best of our knowledge, these $\mathcal{W}_k^{(n)}$ algebras have not appeared in the literature before. They contain the $k$ NASS state with $n$ components $\textrm{SU}(n+1)_k /\textrm{U}(1)^{n}$ as a special case. They are relatively exotic, and even for $k=2$  they already contain $n(n-1)/2$ spin $2$ chiral fields, as can be seen from the character of the identity of the simplest case, namely the $k=2$ NASS state 
\begin{equation}
\textrm{Tr} \left( q^{L_0} \right) = \sum_{m_1, \dots, m_{n} \geq 0} \frac{q^{2 m_i^2 - 2 m_i m_{i+1}}}{\prod_i (q)_{2m_i}} = 1 + \frac{n(n-1)}{2}q^2 +O(q^3)
\end{equation}
It is therefore natural to ask whether these CFTs contain several stress-energy tensors, and can be factorized into several simpler CFTs. This factorization is supported by the naive coset decomposition: 
\begin{equation}
\frac{\textrm{SU}(k)_l \otimes (\textrm{SU}(k)_1)^{n}}{\textrm{SU}(k)_{n+l}}  \simeq  \frac{\textrm{SU}(k)_l \otimes \textrm{SU}(k)_1}{\textrm{SU}(k)_{l+1}} \otimes  \frac{\textrm{SU}(k)_{l+1} \otimes \textrm{SU}(k)_1}{\textrm{SU}(k)_{l+2}} \otimes \cdots \otimes \frac{\textrm{SU}(k)_{l+n-1} \otimes \textrm{SU}(k)_1}{\textrm{SU}(k)_{l+n}} \label{pseudo_factorization}
\end{equation}
However this question has already been addressed in a simple case, namely $n=k=r=2$ in \cite{Grosfeld_Schoutens}, and it turns out not to be that simple. In that case the factorization would amount to
\begin{equation}
\frac{\textrm{SU}(3)_2}{\textrm{U}(1)^2} \simeq  \frac{\textrm{SU}(2)_1 \otimes \textrm{SU}(2)_1}{\textrm{SU}(2)_{2}} \otimes  \frac{\textrm{SU}(2)_{2} \otimes \textrm{SU}(2)_1}{\textrm{SU}(2)_{3}} 
\end{equation}
 In \cite{Grosfeld_Schoutens} it was found that these CFTs are indeed related, but the NASS theory $\textrm{SU}(3)_2 /\textrm{U}(1)^2$ does not exactly factorizes into  Ising  ($c=1/2$) $\otimes$  Tri-critical Ising ($c=7/10$).   In particular the algebras of these two CFTs have to be extended by a fermion parity operator, effectively doubling their Ramond sectors. Moreover some selection rules have to be imposed to respect this extra fermion parity \cite{Grosfeld_Schoutens}. As we were finishing this work, it was brought to our attention that another group \cite{Davenport_Ardonne_Simon} has been studying the spin-Gaffnian state ($n=k=2,r=3$). They have found a (semi-direct) product of minimal models, confirming our ansatz for the underlying CFT as well as the pseudo-factorization \eqref{pseudo_factorization}
 \begin{equation}
 \frac{\textrm{SU}(2)_{-1/2} \otimes (\textrm{SU}(2)_1)^{2}}{\textrm{SU}(2)_{3/2}} \simeq \frac{\textrm{SU}(2)_{-1/2} \otimes \textrm{SU}(2)_1}{\textrm{SU}(2)_{1/2}} \otimes  \frac{\textrm{SU}(2)_{1/2} \otimes \textrm{SU}(2)_1}{\textrm{SU}(2)_{3/2}} = M(3,5) \otimes M(5,7)
\end{equation}
We conjecture that the CFT underlying the $(k,r)$ Jack state for particles with $n$ internal states is $\mathcal{W}^{(n)}_k(p=k+1,q=k+1 +n(r-1))$. At this point the identification of the underlying algebra as being $\mathcal{W}^{(n)}_k$ is quite speculative. However in the following we give some strong evidence in favor of it: first, this holds true for the spin-polarized case $n=1$, but this is rather trivial. More convincingly, the central charge we extracted from the Jack wavefunctions matches with that of  $\mathcal{W}^{(2)}_k(p=k+1,q=k+1 +2(r-1))$. In the next paragraph we use this conjecture to predict the quasihole conformal dimension, and we recover the NASS quasihole dimensions.

\subsection{Quasihole operators and duality}

For the spin polarized case, it was found in \cite{RB2} that there is a dual action of the CS Hamiltonian on the quasihole coordinates of the $(k,r)$ FQH wavefunctions. This structure was inherited from the CFT, where the electron and quasihole fields were dual from one another. This duality consists in interchanging the roles of $p$ and $q$ in $\mathcal{W}^{(1)}_k(p,q)$, and is very sensitive to the choices of $(p,q)=(k+1,k+r)$ corresponding to the spin-polarized Jack state. On the other hand, this duality comes from the integrable structure of the Calogero-Sutherland model, and should still hold for the case $n\geq 1$. There this duality amounts to exchange $p=k+1$ and $q=k+1+n(r-1)$ in $\mathcal{W}^{(n)}_k(p,q)$. Equivalently one can think of changing the value of $r \rightarrow \tilde{r}$ so that $(k+1,k+1+n(\tilde{r}-1))$ is proportional to $(k+1+n(r-1),k+1)$. The dual value $\tilde{r}$ is then:
\begin{align}
(k+1+n(r-1)) (k+1+n(\tilde{r}-1)) = (k-1)^2 \label{r_tilde}
\end{align}
Under the transformation $r \rightarrow \tilde{r}$ the central charge remains unchanged, but the primary field representing the electron transforms into another primary field, with a different conformal dimension:
\begin{align}
\frac{r}{2} \frac{k-1}{k} \rightarrow \frac{\tilde{r}}{2} \frac{k-1}{k} \label{dual_dimension}
\end{align}
In the spin polarized case ($n=1$) this other primary field is precisely the one representing the elementary quasi-hole \cite{RB2}. Assuming this duality relation to hold for generic $n$, and plugging \eqref{r_tilde} in \eqref{dual_dimension}, the quasihole conformal dimensions has to be:
\begin{equation}
\Delta_{\sigma} = \frac{k-1}{2k} \frac{2+2 k-n+(n-1-k) r}{k+1+n (r-1)}
\end{equation}
As a trivial check we recover the conformal weight of $\Delta_{\sigma} = \frac{k-1}{2k} \frac{1 + 2 k - k r}{k + r}$ for the spin polarized case ($n=1$). More interestingly for $r=2$ we recover the $\textrm{SU}(n+1)_k$ NASS quasihole dimension:
\begin{equation}
\Delta_{\sigma} = \frac{k-1}{2k} \frac{n}{k+n+1}
\end{equation}
As a bonus result we find that quasihole correlators are eigenstates of the spin Laplace-Beltrami operator for the following dual value of the coupling:
\begin{equation}
\tilde{\alpha} = n - \alpha
\end{equation}
and $\alpha = - (k+1)/(r-1)$. This implies also that $4$ point functions of $\sigma$'s are simply obtained from replacing $r$ by $\tilde{r}$ in those of the $\Psi$'s. For instance: 
 \begin{equation}
 \langle \sigma^{(\uparrow)}_{-1}(0)\sigma^{(\uparrow)}_1(z)\sigma^{(\uparrow)}_{1}(1)\sigma^{(\uparrow)}_{-1}(\infty)\rangle z^{\tilde{r}\frac{(k-1)}{k}}(1-z)^{\frac{\tilde{r}}{k}} = {}_2F_{1}\left[-\tilde{r},\frac{1-\tilde{r}}{k+1};k\frac{1-\tilde{r}}{k+1};z\right]
\end{equation}
where $\tilde{r} = \frac{2+2 k-n+(n-1-k) r}{k+1+n (r-1)}$. At this point comes a highly non trivial check:  for $(k,r,n)=(2,2,2)$ this quasihole $4$ point function reproduces exactly the results (B.10) and (B.11) of Ref. \cite{Ardonne_Schoutens}. With several strong checks along the way, the conjecture that the CFT underlying the $(k,r)$ Jack state for particles with $n$ internal states is $\mathcal{W}^{(n)}_k(p=k+1,q=k+1 +n(r-1))$ proves to be robust.

\section{Conclusion}

We have generalized the bosonic spin-polarized Jack polynomial FQH states to include spin degrees of freedom. The new polynomials, called Jack polynomials with a prescribed symmetry,  represent spin-singlet states with several special properties. They are eigenstates of a generalized spin Laplace-Beltrami operator,  exhibit clustering properties and a generalized squeezing structure from a root partition that satisfies a generalized $(k,r)$ pauli principle. Our formalism includes, as special cases, the Halperin and the NASS states. We then presented a conjecture for the CFT that describes the non-symmetric Jack polynomials and substantiated it by matching the CFT central charge and quasihole scaling dimension with those obtained from an explicit calculation using the polynomial wavefunctions.  Several things remain to be done: first, a proof of our conjecture is desirable. Second,  the coefficients of each non-interacting many-body state should be obtainable through a recursion relation in the same spirit as for the spin-polarized case. Third, the numerical implementation of such a relation should increase the efficiency of existing spin-unpolarized FQH codes. Relating our construction to the usual squeezing structure but from several root partitions presented in \cite{regnaultardonne} is also desirable.

{\it Acknowledgements}: The authors thanks N.~Regnault for very helpful discussions. B.E. aslo acknowledges conversations with Vl.~Dotsenko, V.~Pasquier, R.~Santachiara, K.~Schoutens and D.~Serban.  BAB thanks R. Thomale for discussions. BAB was supported by Princeton Startup Funds, Sloan Foundation, NSF DMR-095242,  MRSEC grant at Princeton University, NSF DMR-0819860, and by ONR N00014-11-1-0635 grant. BE was supported by the foundation FOM of The Netherlands.

\appendix

\section{Halperin as an eigenstate of the spin Laplace-Beltrami operator}
\label{Halperin_Proof}

In this appendix we show that the bosonic spin-singlet Halperin ground state
\begin{equation}
\Phi = \prod_{i<j}^N (z_i- z_j)^r (w_i-w_j)^r \prod_{i,j}^N(z_i- w_j)^{r-1}
\end{equation}
is an eigenstate of the spin Laplace-Beltrami operator \eqref{sLB}. We work with $N^{\uparrow} = N^{\downarrow} = N $  particles (please keep in mind the change of notation, $N$ is not $N^{\uparrow} + N^{\downarrow}$ as in the rest of the paper). To start with, let us have a closer look at the exchange term of the sLB operator:
\begin{equation}
\Phi^{-1}(z_i,w_j) \sum_{i,j} \left(1- K_{z_i,w_j}\right)\frac{z_i w_j}{(z_i-w_j)^2} \Phi(z_i,w_j)
\end{equation}
It is straightforward to compute the explicit action on the Halperin wavefunction:
\begin{equation}
\Phi^{-1} \sum_{i,j} \left(1- K_{z_i,w_j}\right)\frac{z_i w_j}{(z_i-w_j)^2} \Phi = \sum_{i,j} \frac{z_i w_j}{(z_i-w_j)^2} \left( 1 + \prod_{k \neq i} \left( \frac{w_j-z_k}{z_i-z_k} \right)  \prod_{l \neq j} \left( \frac{z_i-w_l}{w_j-w_l} \right) \right)
\end{equation}
This expression is $\mathfrak{S}_N \times \mathfrak{S}_N$ symmetric, so there can't be poles when $z_i \to z_j $ or $w_i \to w_j$, and one finds:
 \begin{eqnarray}
 \sum_{i,j} \left(1- K_{z_i,w_j}\right)\frac{z_i w_j}{(z_i-w_j)^2} \Phi & = & \left [ 2 \sum_{i,j} \frac{z_i w_j}{(z_i-w_j)^2} - \frac{N(N-1)(2N-1)}{6} \right] \Phi
\nonumber \\  & + &  \frac{1}{2} \left[  \sum_{i, j\neq l} \frac{z_i^2}{(z_i-w_j)(z_i-w_l)} +  \sum_{i, j\neq l} \frac{w_i^2}{(z_j-w_i)(z_l-w_i)}  \right]  \Phi \label{exchange_Halperin}
\end{eqnarray}
To go further, notice that the Halperin wavefunction is annihilated by
\begin{equation}
\left(z_i \frac{\partial}{\partial z_i} - r \sum_{j\ne i} \frac{z_i}{z_i-z_j} - (r-1)\sum_k \frac{z_i}{z_i - w_k}\right)\Phi^{(r,r,r-1)} =0
\end{equation}
from which one can build a second order operator $D_i$ that also annihilates it
\begin{equation}
D_i (z) = \left(z_i \frac{\partial}{\partial z_i} + \sum_{j\ne i} \frac{z_i}{z_i-z_j} \right)\left(z_i \frac{\partial}{\partial z_i} - r\sum_{j\ne i} \frac{z_i}{z_i-z_j} - (r-1)\sum_k \frac{z_i}{z_i - w_k}\right) 
\end{equation}
And similarly for the operator $D_i (w)$ in terms of $w$ (just take in the above $z \leftrightarrows w$ ). Now by summing both operators over $i$ and adding them together, we get after some algebra :
\begin{equation}
H \Phi = E_{r}\Phi
\end{equation}
where the operator $H$ is
\begin{eqnarray}
H  & = & \sum_{i=1}^N \left[ \left( z_i \frac{\partial}{\partial z_i} \right)^2 + \left( w_i \frac{\partial}{\partial w_i} \right)^2 \right] - \frac{r-1}{2} \sum_{i,j; i\neq j} \left[ \frac{z_i+z_j}{(z_i-z_j) }z_i \frac{\partial}{\partial z_i} + \frac{w_i+w_j}{(w_i-w_j) }w_j\frac{\partial}{\partial w_i} \right]  \nonumber \\
& - &   \frac{r-1}{2} \sum_{i,j}  \left[ \frac{z_i+w_j}{(z_i-w_j) } \left( z_i \frac{\partial}{\partial z_i} -w_j\frac{\partial}{\partial w_i}\right) \right] \nonumber \\ 
& + &   (r-1)  \left(2 \sum_{i,j} \frac{z_i w_j}{(z_i-w_j)^2} +  \frac{1}{2} \sum_{k,l,i;k\neq l} \left[ \frac{z_i^2}{(z_i-w_l)(z_i-w_k)} + \frac{w_i^2}{(w_i-z_k)(w_i-z_l)} \right]\right)
\end{eqnarray}
and the eigenvalue is
\begin{eqnarray}
\epsilon_{r}&  = & \frac{1}{6} N \left(4 N^2 r (-2+3 r)+r (-1+3 r)+3 N \left(1+r-4 r^2\right)\right)
\end{eqnarray}
The last line of the operator $H$ is nothing but the exchange term \eqref{exchange_Halperin}, which finishes the proof that the Halperin wavefunction is an eigenstate of the Laplace-Beltrami operator for $\alpha = - 2/(r-1)$.
\begin{align}
H_{sLB}^{(\alpha= - \frac{2}{r-1})} \Phi = E_r \phi
\end{align}
Moreover the energy
\begin{align}
E_r = \frac{1}{6} N \left(1-2 r+6 N (1-2 r) r+3 r^2+2 N^2 \left(1-5 r+6 r^2\right)\right)
\end{align}
is of the form \eqref{energy} for the following  root partition (in orbital occupation)
\begin{equation}
\left(1,0^{r-1},1,0^r,1,0^{r-1},1,0^r, \cdots, 1,0^{r-1},1 \right)
\end{equation}

\section{Four point functions from the spin Laplace-Beltrami operator}
\label{hypergeometric}

In this appendix we show how to compute the following four-point functions \eqref{hyper1} and \eqref{hyper2}.
\begin{align}
 \langle \Psi^{(\uparrow)}_{-1}(0)\Psi^{(\uparrow)}_1(z)\Psi^{(\uparrow)}_{1}(1)\Psi^{(\uparrow)}_{-1}(\infty)\rangle z^{r\frac{(k-1)}{k}}(1-z)^{\frac{r}{k}}  & = {}_2F_{1}\left[-r,\frac{1-r}{k+1};k\frac{1-r}{k+1};z\right] \\
 \langle \Psi^{(\uparrow)}_{-1}(0)\Psi^{(\uparrow)}_1(z)\Psi^{(\downarrow)}_{1}(1)\Psi^{(\downarrow)}_{-1}(\infty)\rangle z^{r\frac{(k-1)}{k}}(1-z)^{\frac{r-1}{k}}  &= {}_2F_{1}\left[1-r,\frac{1-r}{k+1};k\frac{1-r}{k+1};z\right]
\end{align}
under the assumption that any correlation functions of the form
\begin{align}
\langle \Phi_1(0) \Psi^{(\uparrow)}_1(z_1) \cdots \Psi^{(\uparrow)}_1(z_{N^{\uparrow}}) \Psi^{(\downarrow)}_1(w_1) \cdots \Psi^{(\downarrow)}_1(w_{N^{\downarrow}}) \Phi_{2}(\infty)\rangle \prod_{i<j} z_{ij}^{r/k}  \prod_{i<j} w_{ij}^{r/k}   \prod_{i,j} (z_i-w_j)^{(r-1)/k} 
\end{align}
is an eigenstate of the sLB operator \eqref{sLB}, for any primary fields $\Phi_1$ at the origin and $\Phi_2$ at infinity. On the sphere, these extra fields can be thought of as quasi-hole at the poles, since they modify the vanishing properties of the wavefunction at these points. In the spin polarized case it is shown in \cite{EPSS} that even with such insertions, theses correlation functions  are eigenstate of the Laplace-Beltrami operator.

Let us first start with the first correlation function, as it is simpler. The following correlator
\begin{align}
F_1(z,w) =  \langle \Psi^{(\uparrow)}_{-1}(0)\Psi^{(\uparrow)}_1(z)\Psi^{(\uparrow)}_{1}(w)\Psi^{(\uparrow)}_{-1}(\infty)\rangle z^{r\frac{(k-1)}{k}}w^{r\frac{(k-1)}{k}}(w-z)^{\frac{r}{k}}
 \end{align}
is nothing but a $(k,r)$ wavefunction for two up-spins electrons at positions $z$ and $w$ on the sphere, with non trivial boundaries (quasi-holes) at the poles. By construction, it is an eigenstate of sLB, and its dressed partition can be inferred from the vanishing properties at the origin. The partition is $\lambda = (r,0)$ and the spin dressing is $\sigma = (\uparrow,\uparrow)$. From \eqref{energy}  this means that the eigenvalue is $ r (1 + k r)/(1 + k)$. Our function obeys the following partial differential equation:
\begin{align}
\left[ \left( z \partial_z \right)^2 + \left( w\partial_w \right)^2 - \frac{r-1}{k+1}  \frac{z+w}{z-w} (z \partial_z  - w\partial_w) + 2 \frac{r-1}{k+1} \frac{zw}{z-w^2} (1 -  K_{z,w}) \right] F_1(z,w)  = r \frac{1 + k r}{1 + k} F_1(z,w) \label{PDE1}
\end{align}
Moreover this wavefunction is symmetric under $z \leftrightarrow w$, so the exchange term $(1-K_{z,w})$ vanishes identically. Finally this wavefunction is an homogeneous polynomial of degree $r$. Decomposing $F_1(z,w) = w^r P_1(z/w)$,  the two variables partial differential equation \eqref{PDE1} reduces to a one variable differential equation for $P_1$. It turns out to be a simple hypergeometric differential equation, with a unique polynomial solution:
\begin{align}
 \langle \Psi^{(\uparrow)}_{-1}(0)\Psi^{(\uparrow)}_1(z)\Psi^{(\uparrow)}_{1}(1)\Psi^{(\uparrow)}_{-1}(\infty)\rangle z^{r\frac{(k-1)}{k}}(1-z)^{\frac{r}{k}}  & = {}_2F_{1}\left[-r,\frac{1-r}{k+1};k\frac{1-r}{k+1};z\right] 
\end{align}
The second correlator 
\begin{align}
F_2(z,w) =   \langle \Psi^{(\uparrow)}_{-1}(0)\Psi^{(\uparrow)}_1(z)\Psi^{(\downarrow)}_{1}(w)\Psi^{(\downarrow)}_{-1}(\infty)\rangle z^{r\frac{(k-1)}{k}}(w-z)^{\frac{r-1}{k}} w^{(r-1)\frac{k-1}{k}}
 \end{align}
is also an eigenstate of the sLB operator, with eigenvalue $k (r-1)^2/(1+k)$ coming from the partition $\lambda = (r-1,0)$ and spin dressing $\sigma = (\uparrow,\downarrow)$. However the situation is slightly more involved, as it not symmetric under $z \leftrightarrow w$. This issue illustrates the degeneracy of the sLB operator. For a given eigenvalue, there are two solutions of the sLB operator in two variables: a symmetric one and an antisymmetric one. The wavefunction $F_2(z,w)$ is a linear combination of these two solutions $F_2(z,w) = aS(z,w) + bA(z,w)$, for which the partial differential equation reads as:
\begin{align}
\left[ \left( z \partial_z \right)^2 + \left( w\partial_w \right)^2 - \frac{r-1}{k+1}  \frac{z+w}{z-w} (z \partial_z  - w\partial_w) \right] S(z,w)  = r \frac{(r-1)^2}{1 + k} S(z,w) \label{PDE2.1} \\
\left[ \left( z \partial_z \right)^2 + \left( w\partial_w \right)^2 - \frac{r-1}{k+1}  \frac{z+w}{z-w} (z \partial_z  - w\partial_w) + 4 \frac{r-1}{k+1} \frac{zw}{z-w^2}  \right] A(z,w)  = r \frac{(r-1)^2}{1 + k} A(z,w) \label{PDE2.2}
\end{align}
Once again these can be turned into one-variable differential equations using the homogeneity of the symmetric and antisymmetric part, whose polynomial solution turn out are unique:
\begin{align}
S(z,1) = {}_2F_{1}\left[1-r,\frac{1-r}{k+1};1+k\frac{1-r}{k+1};z\right]  \\ A(z,1) = (1-z) {}_2F_{1}\left[2-r,1+\frac{1-r}{k+1};1+k\frac{1-r}{k+1};z\right] 
\end{align}
Therefore the correlation function we  want to compute is of the form
\begin{eqnarray}
 F_2(z,1) = a \, {}_2F_{1}\left[1-r,\frac{1-r}{k+1};1+k\frac{1-r}{k+1};z\right]   + b \,  {}_2F_{1}\left[1-r,\frac{1-r}{k+1};k\frac{1-r}{k+1};z\right] 
 \end{eqnarray}
 where we used one of Gauss' contiguous relations to express $ (1-z) {}_2F_{1}\left[2-r,1+\frac{1-r}{k+1};1+k\frac{1-r}{k+1};z\right] $ as a linear combination of ${}_2F_{1}\left[1-r,\frac{1-r}{k+1};1+k\frac{1-r}{k+1};z\right]$ and ${}_2F_{1}\left[1-r,\frac{1-r}{k+1};k\frac{1-r}{k+1};z\right] $. The coefficients $a,b$ have to be determined. Using the fact that there is no chiral field of dimension $1$, the expansion around $0$ of this function has to be $1  - \frac{r-1}{k} z + O(z^2)$. This forces $a=0$ and $b=1$.
\begin{equation}
 \langle \Psi^{(\uparrow)}_{-1}(0)\Psi^{(\uparrow)}_1(z)\Psi^{(\downarrow)}_{1}(1)\Psi^{(\downarrow)}_{-1}(\infty)\rangle z^{r\frac{(k-1)}{k}}(1-z)^{\frac{r-1}{k}}  = {}_2F_{1}\left[1-r,\frac{1-r}{k+1};k\frac{1-r}{k+1};z\right]
\end{equation}

\end{document}